# Cooperative Caching Framework for Mobile Cloud Computing

By Preetha Theresa Joy & K. Poulose Jacob

Cochin University of Science and Technology, India

*Abstract -* Due to the advancement in mobile devices and wireless networks mobile cloud computing, which combines mobile computing and cloud computing has gained momentum since 2009. The characteristics of mobile devices and wireless network makes the implementation of mobile cloud computing more complicated than for fixed clouds. This section lists some of the major issues in Mobile Cloud Computing. One of the key issues in mobile cloud computing is the end to end delay in servicing a request. Data caching is one of the techniques widely used in wired and wireless networks to improve data access efficiency. In this paper we explore the possibility of a cooperative caching approach to enhance data access efficiency in mobile cloud computing. The proposed approach is based on cloudlets, one of the architecture designed for mobile cloud computing.

*GJCST-E Classification :* C.2.4

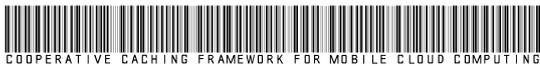

Strictly as per the compliance and regulations of:

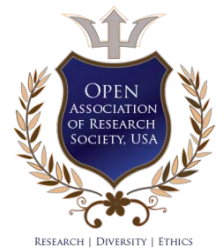

# Cooperative Caching Framework for Mobile Cloud Computing

Preetha Theresa Joy [α] & K. Poulose Jacob [σ]

*Abstract -* Due to the advancement in mobile devices and wireless networks mobile cloud computing, which combines mobile computing and cloud computing has gained momentum since 2009. The characteristics of mobile devices and wireless network makes the implementation of mobile cloud computing more complicated than for fixed clouds. This section lists some of the major issues in Mobile Cloud Computing. One of the key issues in mobile cloud computing is the end to end delay in servicing a request. Data caching is one of the techniques widely used in wired and wireless networks to improve data access efficiency. In this paper we explore the possibility of a cooperative caching approach to enhance data access efficiency in mobile cloud computing. The proposed approach is based on cloudlets, one of the architecture designed for mobile cloud computing.

## I. Introduction

Mobile cloud computing has received large interest recently as it allows storage and processing of data outside the mobile device. It has a growing popularity due to the proliferation of smart phones which act as mini PCs. The limitations of the mobile device such as smaller size, low battery life and other features can be overcome by offloading the processing and storage to a cloud. The offloading can happen to a remote data center, nearby computer or cluster of computers, or even to nearby mobile devices. Cloud computing is a frame work for sharing resources, information and software capabilities to different mobile devices. The resources will be available on the cloud and can be shared by the devices on demand. In mobile cloud computing environment the client can use the cloud to back up data in the mobile devices. Generally, there are two approaches to realize mobile cloud computing namely General Purpose Mobile Cloud Computing (GPMCC) and an Application Specific Mobile Cloud Computing (ASMCC) [5]. GPMCC is utilizing the internet by the mobile devices to use the computing resources of remote computers without any applications specifically developed for this purpose. In ASMCC, specific applications are developed for mobile devices to use the cloud computing facility. Mobile Service Clouds proposed in [1] is a cloud service which uses ASMCC approach for the deployment of autonomic communication services. In [4], mobile cloud computing is broadly classified in into two, those which use mobile devices as thin clients, offloading computation to cloud resources on the internet and the one using mobile devices as computational and storage nodes as a part of cloud computing infrastructure.

Although mobile devices have improved much in processing speed, memory and operating systems, they still have some serious drawbacks. The major challenge for a mobile device in cloud computing is the data transfer bottle neck. Battery is the major source of energy for these devices and the development of battery technology has not been able to match the power requirements of increasing resource demand. The average time between charges for mobile phone users is likely to fall by 4.8% per year in the near future [2]. As the cloud grows in popularity and size, infrastructure scalability becomes an issue. Without scalability solution, the growth will result in excessively high network load and unacceptable service response time.

Data caching is widely used in wired and wireless networks to improve data access efficiency, by reducing the waiting time or latency experienced by the end users. A cache is a temporary storage of data likely to be used again. Caching succeeds in the area of computing because access patterns in typical computer applications exhibits locality of reference [3]. Caching is effective in reducing bandwidth demand and network latencies. In wireless mobile network, holding frequently accessed data items in a mobile node's local storage can reduce network traffic, response time and server load. To have the full benefits of caching, the neighbor nodes can cooperate and serve each other's misses, thus further reducing the wireless traffic. This process is called cooperative caching. Since the nodes can make use of the objects stored in another node's cache the effective cache size is increased. In this paper we discuss a cooperative cache based data access frame work for mobile cloud computing. The proposed approach uses the cloudlet architecture presented by M. Satyanarayanan [7].

The rest of the paper is organized as follows. Section 2 overviews the mobile cloud architecture, section 3 briefs the applications of mobile cloud computing, section 4 presents the limitations of mobile cloud computing, section 5 explains the cloudlet architecture, section 6 details the proposed cooperative cache architecture and section 7 discusses the

*Author α σ :* Cochin University of Science and Technology, India.
*E-mail :* Preetha@mec.ac.in

possibility of exploring caches in other mobile cloud architecture, section 8 concludes the paper.

## II. Mobile Cloud Architecture

The general architecture of MCC is shown in Fig 1. The mobile clients are connected to the internet via base stations, access points or by a satellite link. The shared pool of resources in mobile cloud computing are virtualized and assigned to a group of distributed servers managed by the cloud services. The cloud services are generally classified based on a layered concept [6].The frame work is divided in to three layers, Infrastructure as a Service ( IaaS ), Platform as Service (PaaS) and Software as a Service (SaaS).

Infrastructure as a Service (IaaS) : IaaS includes the resources of computing and storage. It provides storage, hardware, servers and networking components to the user. The examples of IaaS are Elastic Cloud of Amazon and S3 (Simple Storage Service).

Platform as a Service (PaaS): Paas provides an environment of parallel programming design, testing and deploying custom applications. The typical services are Google App engine and Amazon Map Reduce/Simple Storage Service.

Software as Service (SaaS): SaaS provides some software and applications which the users can access via Internet and is paid according to the usage. Google online office is an example for SaaS.

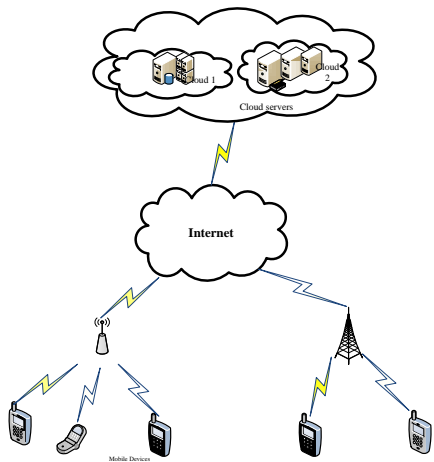

*Figure 1 :* Mobile Cloud Computing Architecture

## III. Applications of Mobile Cloud Computing

Mobile cloud applications move the resource intensive applications and storage away from mobile phones so that the mobile applications are not constrained to certain kind of mobile devices. This helps to overcome the storage capacity and computing power constraints of mobile devices. Mobile cloud computing paradigm is an attractive option for many areas like business, mobile image processing and for computing intensive applications like speech recognition, machine learning augmented reality etc. The typical applications of mobile cloud computing includes Mobile Commerce, Mobile Learning, Mobile Healthcare, Mobile Gaming and other practical applications like social networking, showing maps, storing images and video[6].

## IV. Major Issues in Mobile Cloud Computing

The key elements in a mobile cloud computing approach are: mobile devices, networks through which the devices communicate with the cloud and mobile applications. The major challenge in cloud computing comes from the characters of the first two elements, mobile devices and wireless network .This makes the implementation of mobile cloud computing more complicated than for fixed clouds. This section lists the major issues in Mobile Cloud Computing.

Limitations of the Mobile devices: Compared to personal computers mobile devices have limited storage capacity, poor display, less computational power and energy resource. Although smart phones have improved a lot, they still have battery power constraint.

Network Bandwidth and Latency: As the mobile cloud computing uses wireless networks for data transfer bandwidth is a major issue compared to wired networks which uses a physical connection to ensure bandwidth consistency. Furthermore, the cloud services may be located far away from mobile users, which in turn increase the network latency.

Heterogeneity: Heterogeneity in mobile cloud computing comes from two sources: mobile devices and mobile networks. There is a wide range of mobile devices used by the group of people sharing the network. The operating system and the application software used by these devices vary which cause a major issue in the interoperability of the devices. Another area is the different radio technologies used for accessing the cloud. This will lead to changes in bandwidth and network overlay.

Service Availability: Availability of service is an important issue in mobile cloud computing. Mobile clients may not be able to connect to the cloud due to traffic congestion, network failures and out of signal.

Privacy and Security: Offloading computation and storage to cloud pose security and trust issues. The cloud services are vulnerable and the mobile clients may lose their data if the services fail due to some technical issues.

## V. Cloudlet Architecture

One of the key issues in mobile cloud computing is the end to end delay occurring in servicing a request. Since the cloud services and resources in Internet service provider is distant from the mobile clients network latency is increased. Satyanarayanan (2009) proposed an architecture for mobile cloud computing called Cloudlet, to reduce the bandwidth induced delay between devices and cloud. A cloudlet is a set of computers connected to the internet and is accessed by the nearby mobile devices using a Wi-Fi or WLAN. In this architecture the mobile devices act as a thin client and all the computations occur in the cloudlet. Fig 2 shows cloudlet architecture. The basic idea of this approach is to reduce the distance between the mobile users and cloud services. By this architecture mobile users can access the cloudlet which is one hop away, thus reduces bandwidth utilization and efficiency.

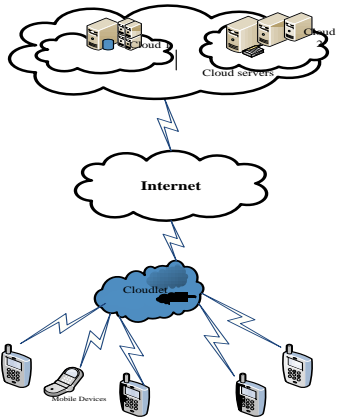

*Figure 2 :* Mobile Cloudlet Architecture

To overcome the challenges of mobile devices, parts of computation and data storage is migrated to resource providers outside the mobile device. Offloading transfers the information needed for processing and storage to a remote server which will complete the computation task and send the results back to the mobile client. The different offloading techniques currently available are client server communication, virtualization and mobile agents [20].

Cloudlet architecture reduces the gap between the mobile devices and remote servers by offloading its workload to a local cloudlet with connectivity to remote cloud servers. Offloading in cloudlet infrastructure is based on a Virtual Machine (VM) technology. To create a transient client software environment cloudlet uses dynamic VM synthesis. In this approach the mobile device transmits a small VM overlay to the cloudlet and applies it to a compatible base VM to generate the Launch VM. Launch VM is the virtual environment temporarily created for a mobile client to execute the task. After execution results are given back to the mobile client. The cloudlet infrastructure is restored to its previous state after each execution.

The performance of cloudlet depends mainly on two factors. Overlay transmission time and overlay synthesis. Overlay transmission time can be improved by using a higher bandwidth wireless network. The different techniques proposed in [7] includes partitioning the overlay in to different chunks that can be executed in parallel, caching and prefetching and pipelining i.e. pipelining the overlay execution with transmission. In this paper we discuss a cooperative cache design for the cloudlet architecture.

## VI. Proposed Approach

Mobile cloud computing has found wide applications in many areas like speech synthesis, natural language processing, image processing, augmented reality, information sharing ,information searching , social networking, etc. While many applications like information sharing or social networking are not dependent on the speed of processing, some computation intensive applications like augmented reality, image processing demand high level of responsiveness. Cooperative caching tries to improve the response time by reducing VM synthesis time by caching previous states. If the users that use cloud services have similar interest, cooperative caching increases the response time considerably. A language translator is an interesting application, which we could look into. This is a useful tool for foreign travelers. Using mobile cloud computing, different words, sentences or paragraphs can be independently processed in the cloud. Commonly used words or sentences will be available in the local cache, which can be accessed faster during subsequent searches, thereby improving the responsiveness of the system.

Cooperative caching consists of multiple distributed caches to improve system response time. Having distributed caches permits a system to deal with concurrent client request as well as sharing contents. We can also reduce response time by concurrently retrieving objects from different cache sites. Concurrent retrieval of objects from different cache sites is beneficial as opposed to the remote cloud server which will result in latency and bandwidth issues.

In cloudlet, when a mobile client requests for a cloud service, the network searches for data in the local cloud. If the service is not available, the users should contact a distinct cloud which involves network transfers and latency. If we are able to cache different VM synthesis states, the users can get the service from cache. When the object is not present in the cache, request is given to the base layer to get the corresponding launch VM. If t the corresponding base h VM is not present, we have to contact distant cloud for the service. Thus data caching also increases battery life in mobile devices by reducing wireless communication.

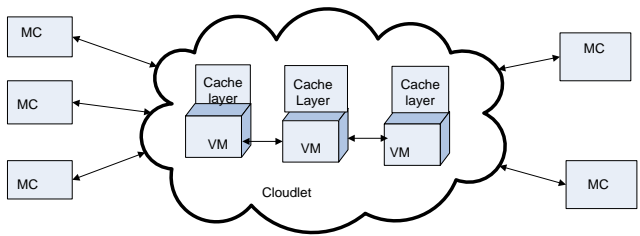

Figure 3 : Cooperative Cache Framework

## VII. Cache Deployment Options

There are two main cache deployment options: those which are deployed in the strategic points in cloudlet based on user access pattern and those which are deployed between the cloudlets. In this paper we consider the first option, deploying cache in different points (virtual machines) in the cloudlet. Fig. 3 shows the cooperative cache frame work for cloudlet architecture. The cloudlet consists of virtual machines which are temporary customization of software environment for each client for their use. The virtual machines separate the transient client software environment from the permanent host software environment. A local cache can reduce virtual machine's synthesis delay by caching virtual machine states that are likely be used again. In a cloudlet we can have more than one virtual machine with a local cache. If we are able to share the cache states, availability and accessibility of different states can be improved. Fig.4 shows the different components of cache layer.

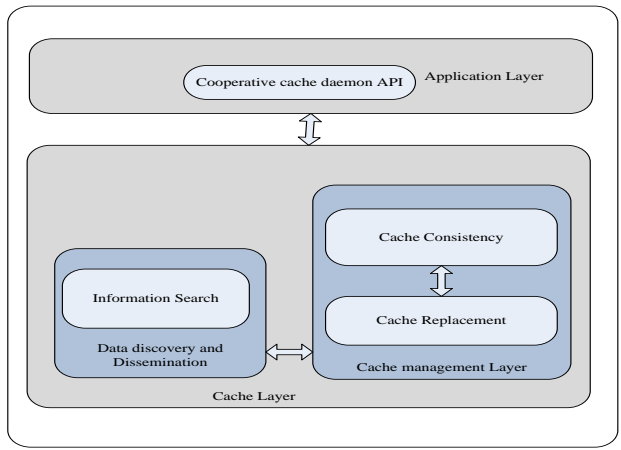

Figure 4 : Different components of cache layer

The cooperative cache daemon API acts as an interface between the application layer and the cache layer. The core system consists of two modules: data discovery and dissemination and the cache management. The information search module in the data discovery and dissemination layer locates and fetches the required object from the cache module. The cache management layer includes the cache replacement and consistency modules. Cache consistency module is designed to be configurable to maintain data synchronization with the original data. The cache replacement module handles the replacement of objects when the cache is full.

The efficiency of a distributed cache depends on three services, discovery, dissemination and delivery of objects. Discovery refers to how the clients locate the cached object. Dissemination is the process of selecting and storing objects in the cache i.e., deciding the objects to be cached, where they are cached and when they are cached. Delivery defines how the objects make their way from the server or cache site to the client. A query based or directory based approach can be used for information discovery. Dissemination may be either client initiated or server initiated. In client initiated dissemination, the client determines what, when and where to cache. The advantage of this scheme is that it automatically adapts to the rapidly changing request pattern. In server initiated dissemination the server chooses the object to be cached. Here the server can maintain a historical data to make the dissemination decision. This approach can provide strong consistency compared to client driven approach. For the proposed approach as the mobile devices act as thin client dissemination decision can be taken by the cloudlet. Another issue we must look into is how to replace the objects from the cache when it is full. A number of cache replacement policies are proposed in literature for wired and wireless networks. The important factors that can influence the replacement process are access probability, recency of request for a data item, number of requests to a data item, size, cost of fetching data from server, modification time, expiration time, distance etc. Based on these parameters we can propose different cache replacement policies suitable for mobile cloud computing. Cooperative caching achieves high hit rates and low response time only if caches are distributed, cache sharing is wide spread and discovery overhead is low.

## VIII. Discussion

Compared with traditional cloud computing mobile cloud computing poses a challenge in the way mobile device access data stored in the cloud. This is due to the inherent challenges of mobile computing such as low bandwidth, mobility, limited storage and battery life. In mobile cloud, user's location is not fixed and the bandwidth provided will also vary for same data access. Having caches closer in proximity to certain group of users is an effective way to reduce average network latencies since there is a correlation between the location of the user and the object requested.

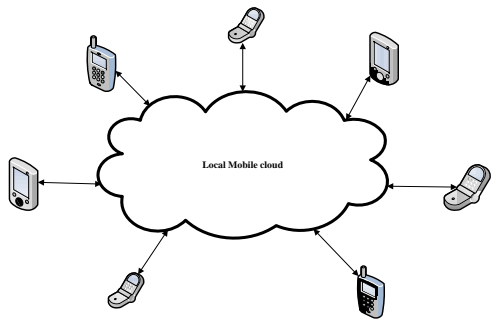

*Figure 5 :* Virtual resource cloud of mobile devices

In this paper we have presented a cooperative caching framework for a cloudlet based mobile cloud computing architecture. The purpose of cloudlet is to reduce the distance between cloud services and mobile devices because when the distance is increased the end to end user delay is also increased, which may not be feasible for some applications. Cloudlet is a middle tier introduced in between the mobile devices and distant cloud services to reduce network latency for the mobile users. By incorporating distributed caches which cooperates each other, we can reduce data traffic and latency considerably. Apart from the caching model discussed for cloudlet architecture, distributed caches can be used in other mobile cloud computing architectures.

Mobicloud [20] is a cloud computing technology proposed for mobile ad hoc networks. In this architecture each mobile node is considered as a service node that can be used as a service provider or service broker depending on its computation and communication capabilities. Every service node is incorporated in to the cloud as a virtualized component and is mirrored in the cloud. As the applications mobile ad hoc networks are targeted for users having similar interest, cooperative caching can be effectively used in this type of networks. Hyrax proposed in [8] considers mobile devices as resource providers of cloud service making up a wireless peer to peer network. The collection of mobile devices that are within the vicinity collaborate each other to form a mobile cloud as shown in Fig 5. In this architecture distributed caches can be used to store data and computational results.

## IX. Conclusions

Mobile cloud computing is a very promising approach for mobile devices. It enables the mobile device to act as thin clients by offloading the computation and processing overhead to cloud servers. Several architectures are proposed in literature for this. The full potential of both cloud computing and mobile applications have not been realized yet. Many deployment challenges have to be addressed before making this a reality. The cooperative caching approach introduced in this paper addresses some key issues in mobile cloud computing .We expect that the challenges posed in implementing our proposal shall be taken up for future studies.

## References Références Referencias

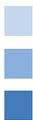